\documentclass[aps,prl,superscriptaddress,twocolumn]{revtex4-2}
\usepackage{graphicx}
\usepackage{color}
\newcommand{\YSO}{Y$_2$SiO$_5$}
\newcommand{\er}{Er$^{3+}$}
\newcommand{\es}{$^4$I$_{13/2}$}
\newcommand{\gs}{$^4$I$_{15/2}$}

\begin{document}

%\preprint{}

\title{Incoherent Measurement of Sub-10 kHz Optical Linewidths}


\author{Félix Montjovet-Basset}
\affiliation{Chimie ParisTech, PSL University, CNRS, Institut de Recherche de Chimie Paris, 75005 Paris, France 
}
\author{Jayash Panigrahi}
\affiliation{Chimie ParisTech, PSL University, CNRS, Institut de Recherche de Chimie Paris, 75005 Paris, France 
}
\author{Diana Serrano}
\affiliation{Chimie ParisTech, PSL University, CNRS, Institut de Recherche de Chimie Paris, 75005 Paris, France 
}
\author{Alban Ferrier}
\affiliation{Chimie ParisTech, PSL University, CNRS, Institut de Recherche de Chimie Paris, 75005 Paris, France 
}
\affiliation{Faculté des Sciences et Ingénierie,  Sorbonne Université, UFR 933, 75005 Paris, France}

\author{Emmanuel Flurin}
\affiliation{Quantronics group, Université Paris-Saclay, CEA, CNRS, SPEC, Gif-sur-Yvette Cedex, France} 

\author{Patrice Bertet}
\affiliation{Quantronics group, Université Paris-Saclay, CEA, CNRS, SPEC, Gif-sur-Yvette Cedex, France} 
\author{Alexey Tiranov}
\affiliation{Chimie ParisTech, PSL University, CNRS, Institut de Recherche de Chimie Paris, 75005 Paris, France 
}
\author{Philippe Goldner}
\email[]{philippe.goldner@chimieparistech.psl.eu}
\affiliation{Chimie ParisTech, PSL University, CNRS, Institut de Recherche de Chimie Paris, 75005 Paris, France
} 

%\thanks{}


\date{\today}

\begin{abstract}

Quantum state lifetimes $T_2$, or equivalently homogeneous linewidths $\Gamma_h = 1/\pi T_2$, are a key parameter for understanding decoherence processes in quantum systems and assessing their potential for applications in quantum technologies. The most common tool for measuring narrow optical homogeneous linewidths, i.e. long $T_2$, is the measurement of coherent photon echo emissions, which however gives very weak signal when the number of emitters is small. This strongly hampers the development of nano-materials, such as those based on rare earth ions, for quantum communication and processing. In this work we propose, and demonstrate in an erbium doped crystal, a measurement of photon echoes based on incoherent fluorescence detection and its variance analysis. It gives access to $T_2$ through a much larger signal than direct photon echo detection, and, importantly, without the need for a highly coherent laser. Our results thus open the way to efficiently assess the properties of a broad range of emitters and materials for applications in quantum nano-photonics.   
\end{abstract}

%\keywords{}

\maketitle



%\section{Introduction}

%Measuring homogeneous linewidths ($\Gamma_h$) is a fundamental tool for understanding decoherence processes in quantum systems. They are related to quantum state lifetimes ($T_2$) by $\Gamma_h = 1/\pi T_2$ and are thus a measure of the time during which quantum states can be used. Of great interest for fundamental studies, this is also a crucial parameter for assessing the usefulness of platforms for applications to quantum technologies.

Measuring quantum state lifetimes $T_2$, or equivalently homogeneous linewidths $\Gamma_h = 1/\pi T_2$,  is a fundamental tool for understanding decoherence processes in quantum systems.  Of great interest for fundamental studies, it also provides the time during which quantum states can be used and is thus  a crucial parameter  for applications to quantum technologies.

In materials such as rare earth (RE) doped crystals at low temperatures, optical $T_2$ can be very long, corresponding to extremely narrow  homogeneous linewidths from a few kHz down to 75 Hz \cite{thielRareearthdopedMaterialsApplications2011}. Combined with long spin coherence times \cite{ledantecTwentythreeMillisecondElectron2021,rancicCoherenceTimeSecond2018,zhongOpticallyAddressableNuclear2015,guptaDualEpitaxialTelecom2023}, this makes them promising systems for quantum memories \cite{busingerNonclassicalCorrelations12502022,lago-riveraTelecomheraldedEntanglementMultimode2021,liuNonlocalPhotonicQuantum2024}, micro-wave to optical transduction \cite{bartholomewOnchipCoherentMicrowavetooptical2020},  or single photon sources for quantum nodes \cite{ourariIndistinguishableTelecomBand2023a,kindemControlSingleshotReadout2020,deshmukhDetectionSingleIons2023,gritschNarrowOpticalTransitions2022}.  Measuring such narrow homogeneous linewidths is however challenging since they are masked by the transitions' inhomogeneous broadening, i.e. the distribution of their frequencies due to varying micro-strain across the crystal structure \cite{goldnerRareEarthDopedCrystals2015}. 

In this case, the preferred technique for evaluating $\Gamma_h$  is the photon echo (PE), an experiment directly derived from the ubiquitous spin or Hahn echo \cite{kurnitObservationPhotonEcho1964}. It
enables monitoring the free evolution of the relative phases between the emitters' superposition states as a function of time. This evolution is directly related to $T_2$ and thus $\Gamma_h$ and numerous studies have shown the efficiency of this technique for measuring $\Gamma_h$ on a broad range of time scales and deciphering complex dynamics in RE doped crystals \cite{goldnerRareEarthDopedCrystals2015, bottgerOpticalDecoherenceSpectral2006, macfarlaneHighresolutionLaserSpectroscopy2002}. 

PE is a spontaneous collective coherent emission occurring after two laser pulses. Its intensity is independent of the coherence lifetime of the laser driving the emitters $T_{2,laser}$ as long as $T_{2,laser}>\tau_{laser}$ where $\tau_{laser}$ is the longest pulse length, typically on the order of a $\mu$s \cite{bottgerOpticalDecoherenceSpectral2006}. This is a key advantage as $T_{2,laser}$ is often orders of magnitude shorter than the emitters' $T_2$. In stark contrast, techniques like spectral hole burning or saturation spectroscopy operate in the opposite domain, i.e. $T_{2,laser}\gg T_2$, and are thus much more demanding and in addition prone to experimental biases such as power broadening \cite{goldnerRareEarthDopedCrystals2015,macfarlaneHighresolutionLaserSpectroscopy2002, weissErbiumDopantsNanophotonic2021}.  The energy released in the coherent PE emission is however only a small fraction of the total energy stored after excitation \cite{allenOpticalResonanceTwolevel1987}. This is not a problem for bulk materials where the collective enhancement, proportional to the square of the number of emitters $N$, is large.  It makes however direct measurements of PE emission very difficult when $N$ becomes small. This is the case for nanostructured materials, like nanoparticles, thin films, and milled or implanted bulk crystals, which nevertheless offer unique possibilities for greatly enhanced light-matter coupling and integration into photonic circuits \cite{ourariIndistinguishableTelecomBand2023a,kindemControlSingleshotReadout2020,deshmukhDetectionSingleIons2023,gritschNarrowOpticalTransitions2022,zhongEmergingRareearthDoped2019, fossatiFrequencyMultiplexedCoherentElectrooptic2020, guptaDualEpitaxialTelecom2023}. To be able to measure PE in such systems, one has to apply an additional laser pulse at the PE emission time in order to convert the emitter's superposition states into excited or ground state populations, detect their fluorescence and in this way retrieve a large fraction of the stored energy \cite{ulanowskiSpectralMultiplexingTelecom2022}. However, the intensity of the incoherent fluorescence signal depends on the relative phase between the laser pulse and the PE, which suggests that a highly coherent laser has to be used. 

\begin{figure*}
\includegraphics[width=\textwidth]{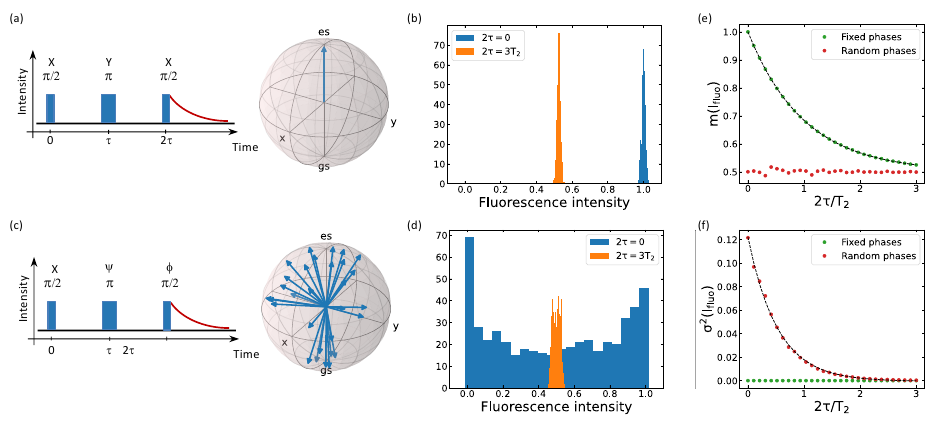}%
\caption{Principle of fluorescence-detected photon echoes (PE). (a) PE sequence with fixed $X-Y-X$ phases for the three laser pulses. Fluorescence (red line) is integrated after the last $\pi/2$ pulse which converts coherences into populations. For evolution times $2\tau\ll T_2$, the final state after the last $\pi/2$ pulse is the excited state (es), as shown in the Bloch sphere. (b) Histogram of the integrated fluorescence intensity ($I_\mathrm{fluo}$) for sequence (a) and different evolution times. A gaussian  noise ($\sigma = 0.01$) has been added to $I_\mathrm{fluo}$  and $T_2\ll T_1$ is assumed.     
(c) Same sequence as in (a) but with randomly varying phases $\psi$ and $\phi$. The final state on the Bloch sphere is a random vector. (d) Histogram of $I_\mathrm{fluo}$ for sequence (c). (e,f) Mean ($m$) and variance ($\sigma^2$) of $I_\mathrm{fluo}$, normalized to $N$ and $N^2$ (see text), for sequences (a), fixed phases,  and (b), random phases. Exponential fits of  $\sigma^2$ and $m$ for (a) and (c) respectively (black  lines) result in the same value of $T_2$ (see text).  \label{fig:1}}
\end{figure*}

In this work, we show that in fact homogeneous optical linewidths can be measured using  detection of an incoherent signal as well as incoherent excitation. Using an \er\ doped \YSO\ single crystal, a material actively investigated for quantum memories and single photon sources \cite{rancicCoherenceTimeSecond2018,ulanowskiSpectralMultiplexingTelecom2022, carSelectiveOpticalAddressing2018},  we recorded  a photon echo decay monitoring fluorescence which gives $T_2 = 37 \pm 2$ $\mu$s or $\Gamma_h = 8.6 \pm 0.4$ kHz. As $T_{2,laser} = 15 \pm 1$ $\mu$s, the laser used was indeed mostly incoherent on the time scale of PE decay. This was achieved by measuring the fluorescence intensity variance instead of mean value \cite{billaudMicrowaveFluorescenceDetection2023} of only $\approx$ 2500 ions. Our results demonstrate a new tool to probe narrow optical homogeneous linewidths of low numbers of emitters. This is especially relevant to the understanding of optical quantum states dynamics in nanostructured materials and to their application to quantum technologies.

The PE echo sequence is depicted in Fig.\ref{fig:1}(a) in the case of well-defined phases $X-Y-X$ for all pulses and an evolution time $2\tau \ll T_2$. For clarity, we also assume $T_2 \ll T_1$, where $T_1$ is the excited state population lifetime.  The first pulse creates superposition states in an inhomogeneously broadened ensemble of emitters and corresponds to a $\pi/2$ rotation along the $X$ axis of the Bloch sphere. After a free evolution time $\tau$, a second pulse rotates the states by an angle $\pi$ along the $Y$ axis and reverses the phase evolution of the emitters' states. After a second period of free evolution, all states are back in phase at time $2\tau$ and align with the $Y$ axis. The coherent echo emission should take place at this time but instead, a third $\pi/2$  pulse along $X$ is applied which sends the emitters to the excited state (Fig. \ref{fig:1}(a)). They afterwards relax by spontaneous emission which is integrated over the excited state population decay. 

When the delay $\tau$ is increased, more and more emitters are randomly perturbed which results in a lower coherence at the echo time and shorter Bloch vector along $Y$. After the last $\pi/2$ pulse the excited state population is also decreasing and ultimately reduces to the center of the Bloch sphere, i.e. equal populations in the excited and ground states. Fig. \ref{fig:1}(b) shows a simulation of the distribution of the integrated fluorescence intensity $I_\mathrm{fluo}$ at $2\tau=0$ and $2\tau=3T_2$, adding noise to take into account detector dark counts for example. 
As expected, $I_\mathrm{fluo}$ mean value decreases as a function of $2\tau$ and its distribution has a constant width.

The sequence shown in Fig. \ref{fig:1}(c) corresponds to the case where the second and third pulses have phases $\psi$ and $\phi$ that randomly vary from shot to shot  with respect to the first pulse phase, arbitrarily chosen as $X$. In this case, the final state after the last $\pi/2$ pulse is a random vector  in the Bloch sphere, and the resulting distribution of $I_\mathrm{fluo}$, given by the projection of the vector on the vertical axis, is shown in Fig. \ref{fig:1}(d). In opposition to the case of fixed phases, the mean fluorescence intensity is constant whereas the distribution width decreases with increasing evolution time. This suggests that the variance of the fluorescence intensity can be used to retrieve the PE decay and thus the coherence lifetime of the emitters. 

Assuming an exponential echo decay of the form $I_{echo} = I_0 \exp[-2(2\tau/T_2)]$,
the integrated fluorescence intensity for sequence (c) reads:
\begin{equation}
    I_\mathrm{fluo} = \frac{N\left[1+\exp(-2\tau/T_2)\cos(\pi-2\psi+\phi)\right ]}{2},
\label{eq:fluo-intens}
\end{equation}
where $N$ is the number of emitters.
For random distributions of the $\phi$ and $\psi$ angles, and in the absence of additional noise, this leads to the mean and variance of $I_\mathrm{fluo}$:
\begin{eqnarray}
    m(I_\mathrm{fluo}) &=& N/2\\
    \sigma^2(I_\mathrm{fluo}) &=& \frac{N^2\exp[-2(2\tau/T_2)]}{8}.\label{eq:fluo-var}
\end{eqnarray}
$T_2$ can thus be determined by recording $\sigma^2(I_\mathrm{fluo})$ even with a laser incoherent over the time range of the PE decay.
It is equivalent to monitoring $I_\mathrm{fluo}$ in the case of 
fixed phases. Indeed, in sequence (a), $\psi = \pi/2$ and $\phi =0$ so that \(m(I_\mathrm{fluo}) = N\left[1+\exp(-2\tau/T_2)\right]/2 \) and  $\sigma^2(I_\mathrm{fluo})=0$. Simulated decays of $m(I_\mathrm{fluo})$ and $\sigma^2(I_\mathrm{fluo})$ in the two  cases are presented in Fig. \ref{fig:1}(e,f). We note that $\sigma^2(I_\mathrm{fluo})$ decays as a function of $2\tau$ as the intensity of the directly detected echo \cite{goldnerRareEarthDopedCrystals2015}.

\begin{figure}
\includegraphics[width=\columnwidth ]{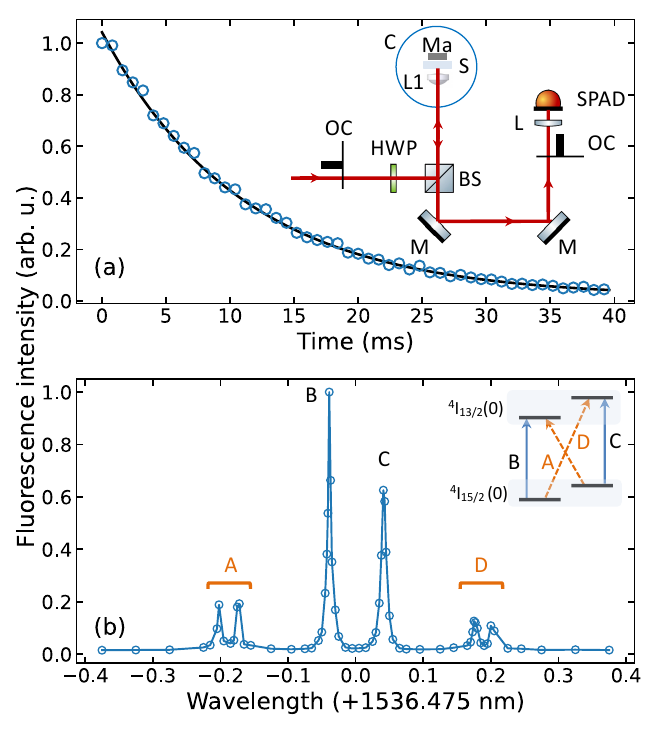}%
\caption{Fluorescence properties of \er\ ions in \YSO\ (site 1) at 3.5 K. (a) Fluorescence decay of the \es $\rightarrow$ \gs\ transition excited at 1536.471 nm (vac.) and exponential fit (black line) giving $T_1 = 11.0 \pm 0.1$ ms. (inset) Schematic of fluorescence microscopy setup. OC: optical chopper, HWP: half-wave plate, BS: beam splitter, L1, L: lenses, S: sample, C: cryostat, M: mirrors, Ma: Magnet, SPAD: single-photon avalanche photodetector. (b) Excitation spectrum of the \gs(0) $\to$ \es(0)\ transition with light polarized along the $b$ axis and magnetic field $\mathbf{B} \parallel D2$. Wavelength in vacuum. The A-D lines are due to Zeeman splittings in the ground and excited states (inset and main text). \label{fig:2}}
\end{figure}

We next experimentally investigated this method using a 10 ppm doped \er:\YSO\ crystal. Part of the setup is shown in the inset of Fig. \ref{fig:2}(a). The sample was cut as a polished slab with a thickness of 500 $\mu$m and cooled to 3.5 K in a closed-cycle cryogenerator, with a permanent magnet producing a field of 200 mT. Excitation was provided by a tunable extended cavity diode laser with no active stabilization. A fluorescence microscopy arrangement was used to excite and collect emission from \er\ ions through a high numerical aperture lens. A single photon InGaAs avalanche photodiode was used for detection and synchronized optical choppers protected the detector from the laser pulses and cut laser stray light during detection. Fluorescence was collected from about 2500  ions and integrated over 40 ms after a delay of 3 ms which was needed to avoid laser leaks due to choppers' jitter. More details can be found in the Supplemental Material (SM).

Fig. \ref{fig:2}(a) shows the fluorescence decay obtained by exciting \er\ ions in site 1 of \YSO\ at 1536.471 nm (vacuum) during 8 $\mu$s.  This wavelength corresponds to the $^4$I$_{15/2}$(0)$\to$$^4$I$_{13/2}$(0) transition between the lowest crystal field levels of the excited and ground multiplets \cite{bottgerSpectroscopyDynamicsEr32006}. Thanks to pulsed excitation and gated detection, fluorescence could be collected on all $^4$I$_{13/2}$(0)$\to$$^4$I$_{15/2}$(0-7) transitions. 
%This increased the signal approximately five-fold compared to the detection of only the $^4$I$_{13/2}$(0)$\to$$^4$I$_{15/2}$(0) coherent transition.  
An exponential fit of the decay gives $T_1=11.0 \pm 0.1$ ms, confirming the origin of the emission \cite{bottgerSpectroscopyDynamicsEr32006} and long enough to allow using optical choppers for strongly suppressing laser light. 

An excitation spectrum was then recorded by integrating fluorescence decays (Fig. \ref{fig:2}(bottom)). It shows four groups of lines that are due to the electronic Zeeman splitting of the $^4$I$_{13/2}$(0) ($g\approx 10.5$) and $^4$I$_{15/2}$(0) ($g\approx 7.0$) levels \cite{sunMagneticTensors4I152008}. Additional lines were observed for the A and D groups and assigned to magnetically non-equivalent sites related by a $C_2$ rotation of the crystal unit cell. From the positions to the lines and known $g$ tensors, the field amplitude was determined at 200 mT and in a direction slightly away from the crystal $D2$ axis. Echo experiments were performed on the B line which has an inhomogeneous linewidth of about 0.7 GHz. 

\begin{figure}
\includegraphics[width=\columnwidth ]{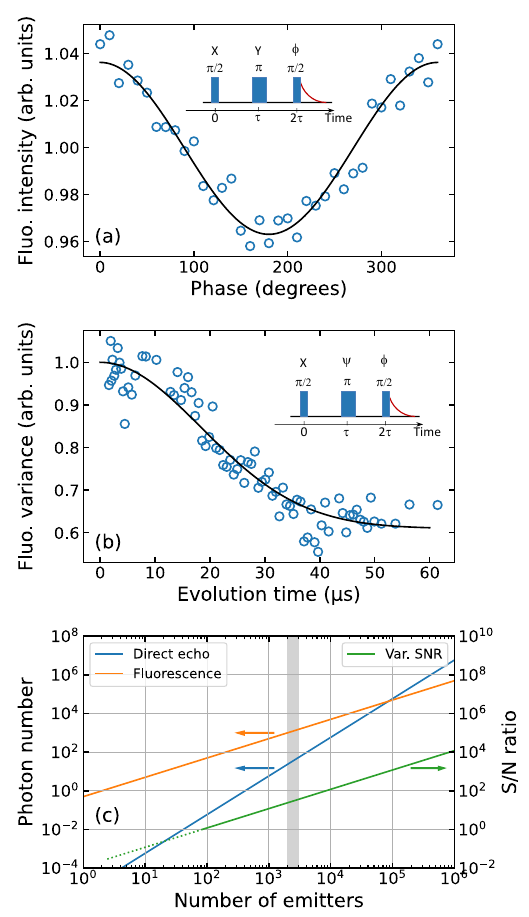}%
\caption{Fluorescence-detected photon echoes. (a) Variation of $m(I_\mathrm{fluo})$ as a function of phase $\phi$ and fit with a cosine function (see text). Inset: pulse sequence, $2\tau =  1.6$ $\mu$s. 
(b)  $\sigma^2(I_\mathrm{fluo})$ as a function of the evolution time $2\tau$ with random phases $\psi$ and $\phi$. Inset: pulse sequence. A fit with a stretched exponential ($\beta=2$) gives $T_2 = 37 \pm 2$ $\mu$s (black line).
(c) Theoretical integrated intensities for fluorescence-detected (orange) and direct (blue) echoes for unit collection efficiencies and this work's experimental parameters. Green line: signal to noise ratio for the fluorescence intensity variance (see text). Gray vertical area: estimated number of ions in (a) and (b). Arrows denote relevant vertical scales.
\label{fig:3}}
\end{figure}

Next, we first used a delay $\tau  = 0.8$ $\mu$s to record fluorescence-detected photon echoes  as depicted in Fig. \ref{fig:1}(a). Indeed, for such short delays, the laser coherence is sufficient to preserve relative phases between the three pulses. A plot of  $m(I_\mathrm{fluo})$ as a function of the last $\pi/2$ pulse phase $\phi$ is shown in Fig. \ref{fig:3}(a). Each data point was averaged over 1000 shots.
A cosine variation was observed in agreement with  Eq. \ref{eq:fluo-intens}, which confirmed that an echo signal was indeed detected by fluorescence. Further proof was given by experiments carried out on a larger sample for which echoes could be directly detected (see SM). 
%Their shapes and timings were found to be in very close agreement with fluorescence-detected ones (see SM).  
Moreover, by repeating the fixed phase experiment for increasing delays $\tau$, we found a laser coherence lifetime $T_\mathrm{2,laser} = 15 \pm 1$ $\mu$s or $\Gamma_\mathrm{laser} = 21 \pm 2 $ kHz (see SM). 
As seen in Fig. \ref{fig:3}(a)), the peak-to-peak variation of $m(I_\mathrm{fluo})$ vs. $\phi$ was about 8\% of the mean value. This is mainly attributed to an inhomogeneous Rabi frequency across the ions because of the small frequency range addressed by the laser (1 MHz) compared to the transition inhomogeneous broadening (0.7 GHz).
%A simple simulation taking only the spectral inhomogeneity into account, modeled as a Gaussian distribution of 1 MHz width, predicts variations of \PG{XX} \%, \PG{close} to the observed value. 

Echo decays were finally recorded using $\sigma^2(I_\mathrm{fluo})$, following the scheme presented in Fig. \ref{fig:1}(c). To avoid different time regimes of laser stability, we randomized the phases of the pulses through the digital waveforms used to create them. The variance was calculated from 2000 shots for each delay $\tau$ at a repetition rate of 10 Hz. 
Fluorescence intensity varied from shot to shot by a few \% for a given  delay, indicating good short-term stability of the experiment and the lack of spectral hole burning. 
%It showed however some slow drift across the duration of the whole measurements, which could reach up to several hours. This behavior is attributed to changes in the laboratory and cryostat temperatures that affect the experiment alignment, and long-term laser frequency drifts. We did not compensate for this in the variance calculations. 
For each delay, about 200 outlier values (larger than 5 standard deviations), attributed to synchronization issues with the choppers and counting system,  were observed and discarded.  

A clear decay of $\sigma^2(I_\mathrm{fluo})$ as a function of $2\tau$ was observed corresponding to a coherence lifetime of $37\pm 2$ $\mu$s according to a stretched exponential fit (Fig. \ref{fig:3}(b)). This is in qualitative agreement with the expected value under our conditions of temperature and magnetic field. The main decoherence process is  attributed to spectral diffusion due to the \er\ spin bath \cite{bottgerOpticalDecoherenceSpectral2006,carSuperhyperfineInducedPhotonecho2020}. The optical setup did not allow to directly record the photon echo because of the several ms delay between the last $\pi/2$ pulse and the opening of the detection window.
% Moreover, as shown below, with 1000 detected ions, we expect an integrated echo signal 7 orders of magnitude smaller than the fluorescence one and thus very likely to be unobservable on our setup. 
We could however confirm that $\sigma^2(I_\mathrm{fluo})$  decay  was identical to the one obtained by direct detection of the echo on the larger sample mentioned above (see SM). 

The results presented above were obtained on a small number of ions. The latter will be typically located within a sub-wavelength volume (a single particle or thin film) or coupled to an optical resonator or waveguide so that the collection efficiencies of the direct and fluorescence-detected echoes are equal. In the following, we consider them equal to 1 for simplicity. The  directly detected integrated echo signal for $N$ ions and $2\tau \ll T_2$ is then given by \cite{billaudMicrowaveFluorescenceDetection2023}:
\begin{equation}
    I_\mathrm{echo} = \frac{N^2 A_{sp}\tau_{echo}}{4}\\
\end{equation}
%where $A_{sp} = 27$ s$^{-1}$ % 54.6 s-1 in Longdell's paper
where $A_{sp}$ is the spontaneous emission rate on the echo transition and $\tau_{echo}$ is the echo pulse length. This formula is valid in the case of low absorption where echo absorption and amplification can be neglected and for $A_{sp}\tau_{echo}\ll 1$. Fluorescence detection gives $I_\mathrm{fluo} = N/2$ and is $\approx 40$ times larger than $I_\mathrm{echo}$ in our experiments ($N=2500$, $A_{sp} = 27$ s$^{-1}$ \cite{bottgerSpectroscopyDynamicsEr32006}, and $\tau_{echo} = 800$ ns) as shown in Fig. \ref{fig:3}(c). The ratio $I_\mathrm{fluo}/I_{echo}$ is proportional to $N$, evidencing the interest of fluorescence detection for small ensembles of emitters. 

The lower bound on $N$ can be evaluated by estimating the noise affecting $\sigma^2(I_\mathrm{fluo})$. For low $N$ it is dominated by the fluorescence intensity shot-noise, which has a variance of $\sigma_{noise}^2(I_\mathrm{fluo}) = N/2$. If $n$ experimental values are used to compute $\sigma^2(I_\mathrm{fluo})$, the standard deviation on $\sigma_{noise}^2(I_\mathrm{fluo})$ is $N/\sqrt{2n}$ for $n, N/2\gg
 1$. The signal-to-noise ratio at $2\tau\ll T_2$ is then proportional to $N$: $R   = k^2N\sqrt{n/32}$, 
where $k$, defined by $\sigma^2(I_\mathrm{fluo})=k^2N^2/8$, takes into account the inhomogeneous Rabi frequency effect.
%for $2\tau \ll T_2$. 

In the experiments presented above, $k = 0.04$ (Fig. \ref{fig:3}(c)) and SM) and $n=1800$, giving a lower bound $N=84$ for $R=1$.   This evaluation assumes that other sources of noise related to excitation and detection are negligible compared to the shot-noise contribution.
%or noise is negligible which could be achieved with a superconducting nanowire system. The intensity noise of our unstabilized laser ($<1$\%) should be negligible compared to the shot-noise contribution, as should be its frequency jitter ($< 200$ kHz over 50 ms), given the short excitation pulses. 
%its frequency jitter ($< 200$ kHz over 50 ms)
To further decrease $N$ or increase $R$,  $k$ could be improved by creating an ensemble of emitters with a linewidth comparable to the excitation bandwidth inside a larger non-absorbing region. This can be done using spectral hole burning techniques and results in much better Rabi frequency homogeneity \cite{guillot-noelHyperfineStructureHyperfine2009}. It is important to note that at least two emitters are required for the method to work since otherwise there is no distinction between decoherence and random laser pulse phases. The condition $T_{2,laser}\gg T_2$ is therefore needed for fluorescence-detected photon echoes on single ions \cite{kindemControlSingleshotReadout2020}.

In conclusion, we have shown that photon echoes can be created and detected using an excitation source and detection that are both incoherent on the time scale of the experiment. This is achieved by converting the echo into populations with an additional laser pulse and measuring the subsequent integrated fluorescence intensity variance instead of its mean value. We demonstrate the validity of the concept using an \er:\YSO\ crystal in which we record an echo decay leading to a coherence lifetime $T_2 = 37 \pm 2$ $\mu$s ($\Gamma_h = 8.6 \pm 0.4$ kHz) whereas $T_\mathrm{2,laser} = 15 \pm 1 $ $\mu$s ($\Gamma_\mathrm{laser} = 21 \pm 2 $ kHz).  Furthermore, these data were acquired using about 2500 ions, which shows the potential of this method to work with small ensembles of emitters, with a shot-noise limited lower limit estimated in our system below 100 ions. This in turn should enable one to fully benefit from the robustness and multi-timescale capability of the photon echo technique to efficiently assess the coherence dynamics of nanostructured materials that find applications in quantum nanophotonics.

\begin{acknowledgments}
This project has received funding from the European Research Council (ERC) under the European Union’s Horizon 2020 research and innovation programme (RareDiamond, grant agreement No 101019234) and the French Agence Nationale de la Recherche under grants ANR-20-CE09-0022 (UltraNanOSpec) and ANR-23-CE47-0011 (MolEQuBe), and the Plan France 2030 project ANR-22-PETQ-0010 (QMemo).
\end{acknowledgments}

\bibliography{references}

\end{document}